\documentclass[11pt]{article}
\pdfoutput=1
\usepackage{epsfig}\parskip 5pt plus 1pt
\usepackage{bbm}
\usepackage{amsmath}
\usepackage{amssymb}
\usepackage{amsfonts}
\usepackage{mathrsfs}
\usepackage{graphicx}
\usepackage{color}
\newcommand{\be}{\begin{equation}}
\newcommand{\ee}{\end{equation}}
\newcommand{\bea}{\begin{eqnarray}}
\newcommand{\eea}{\end{eqnarray}}

\def\gsim{\mathrel{\lower3pt\hbox{$\sim$}}
 \hskip-11.5pt\raise3pt\hbox{$>$}\;}

\def\lsim{\mathrel{\lower3pt\hbox{$\sim$}}
 \hskip-11.5pt\raise3pt\hbox{$<$}\;}

\begin{document}
{\vspace*{-2cm}
\flushright{ULB-TH/09-20}
\vspace{-4mm}
\vskip 1.5cm}
\begin{center}
{\Large \bf
Confined  hidden vector dark matter}
\end{center}
\vskip 0.5cm
\begin{center}

{\large Thomas Hambye and Michel H.G. Tytgat}\footnote{thambye@ulb.ac.be;
mtytgat@ulb.ac.be}
\\
\vskip .5cm
Service de Physique Th\'eorique,\\
\vspace{0.7mm}
Universit\'e Libre de Bruxelles, 1050 Brussels, Belgium\\

\end{center}
\vskip 0.5cm

\begin{abstract}
We argue that the lightest vector bound states of a confining hidden sector communicating with the Standard Model  through the Higgs portal are stable and are viable candidates of dark matter.
The model is based on an $SU(2)$ gauge group with a scalar field in its fundamental representation and the stability of the lightest vector bound state results from the existence of a custodial symmetry.
As the relic density depends essentially on the scale of confinement in the hidden sector, $\Lambda_{HS}$, agreement with WMAP abundance requires $\Lambda_{HS}$ in the $20-120$ TeV range.    
\end{abstract}
\setcounter{footnote}{0}
\vskip2truecm


There are many evidences of the existence of dark matter (DM)\cite{Amsler:2008zz}. The standard explanation is that it is made of a relic stable particle, but the nature of this particle,  the value of the associated energy scale and the origin of its stability are still unclear. 
To explain the latter,  a discrete (or sometime continuous) symmetry is usually invoked, like R-parity for the neutralino DM candidate in supersymmetric models \cite{neutralino}. However, models of dark matter in which this symmetry, rather than put by hand, is the consequence of a gauge symmetry and of the particle content involved, 
or more generally of a fundamental principle, are not many.\footnote{One example of such models arises in specific 
supersymmetric models where R-parity is a remnant group of a gauge symmetry, e.g.~gauged $U(1)_{B-L}$, spontaneously broken at  a higher scale.
Another possibility arises if we assume the existence of 
high dimension $SU(2)_L$ multiplets, a quintuplet or higher for a fermion, or a septuplet or higher for a scalar \cite{Cirelli:2005uq}.
There are also mirror models \cite{foot} and a model in which the (fermion) dark matter candidate is charged under a new $U(1)$ symmetry \cite{Pospelov:2007mp}. }
Recently one simple possibility of this kind has been proposed, by showing that the custodial symmetry which derives from a gauge symmetry could insure the stability of DM \cite{Hambye:2008bq}. The argument supposes the existence of a new  gauge group that may  couple with the SM particles through the Higgs portal, specifically $H^\dagger H \phi^\dagger \phi$ with $H$ the SM Higgs boson doublet and $\phi$ a scalar multiplet of the new gauge group.
To give  mass to the gauge bosons it was assumed in \cite{Hambye:2008bq} that  the  gauge symmetry was spontaneously broken by the vacuum expectation value (vev) of $\phi$. This results in  a multiplet of DM candidates in the form of the massive gauge bosons. Their stability is ensured by the fact that they are the lightest non-singlet particles of a custodial ({\em i.e.}~global) symmetry automatically remaining, even though the gauge symmetry is completely broken.  As for the proton in the Standard Model, such a framework explains the stability of DM particles directly from the low energy content of the model, and does not require to assume any, hard to test, high energy mechanism.\footnote{The proton is an example of stable composite particle. If it had been a fundamental particle destabilizing renormalisable B-violating operators like ${\cal L} \supset  \pi^0 \bar e^c p$ would be allowed. Fortunately it is a bound state of quarks, whose Lagrangian has an accidental global B symmetry \cite{Zee:2003mt}.}

In this short letter we extend the proposal of \cite{Hambye:2008bq} and show that a stable (multiplet of) DM candidates also exists if the model is in a confined instead of the Higgs phase.\footnote{This is presumably to be anticipated, because there are similarities in  the two phases for matter ({\em ie} $\phi$) in the fundamental representation \cite{Fradkin:1978dv}. There is however a fundamental distinction since in the strongly coupled phase, all physical states are bound states rather than elementary particles \cite{'tHooft:1998pk}.} Regarding the stability, our key point is that the custodial symmetry  holds also at the non-perturbative level. The lightest bound state charged under this custodial symmetry may then be a perfectly viable DM candidate with a potentially interesting phenomenology.

For concreteness, we consider the simplest example of such a hidden sector model. It is based on an $SU(2)$ gauge group in the hidden sector, that we denote $SU(2)_{HS}$. To draw an analogy with QCD, we will refer to hidden-color and will envision hidden-color confinement. We introduce also a scalar hidden-color doublet $\phi\equiv(\phi^+ \,\,\phi^0)^T$. All the SM  particles are singlet under $SU(2)_{HS}$. 
The most general renormalisable Lagrangian which can be written is \cite{Hambye:2008bq}
\be
{\cal L}= {\cal L}^{SM} -\frac{1}{4} F^{\mu\nu} \cdot F_{\mu \nu}
+(D_\mu \phi)^\dagger (D^\mu \phi) -\lambda_m \phi^\dagger \phi H^\dagger H-\mu^2_\phi \phi^\dagger \phi -\lambda_\phi (\phi^\dagger \phi)^2 \,,
\label{inputlagr}
\ee
where $D^\mu \phi=\partial^\mu \phi - i\frac{g_\phi}{2} \tau \cdot A^\mu$, with $g_\phi$ the $SU(2)_{HS}$ gauge coupling.  We define the SM Higgs potential as: ${\cal L}^{SM} \supset -\mu^2 H^\dagger H -\lambda (H^\dagger H)^2$ with $H=(H^+,H^0)^T$. Communication between the two sectors is possible only through the Higgs portal term $\phi^\dagger \phi H^\dagger H$, parameterized by the coupling $\lambda_m$, while communication through kinetic mixing, which would destabilize the DM candidate, is forbidden by the non-abelian character of the gauge symmetry.

We now consider the confined phase of this model, {\em ie}  we assume that the perturbative vev, $v_\phi$,
either vanishes or is 
small enough for the running coupling $g_\phi$ to become large at a higher scale $\Lambda_{HS}$, $\Lambda_{HS} >> v_\phi$. 

This model is significantly different from QCD.  There is only one matter field $\phi$ which is scalar and thus no  equivalent of chiral symmetry.  The theory has however an extra, global $SU(2)_{HS}^\prime$ symmetry. 
Just like the Higgs in the SM, the $\phi$ field may be seen as a four real component object, transforming in the real representation of a group $SU(2)_{HS} \otimes SU(2)^\prime_{HS} \sim SO(4)$. Equivalently, the global group $SU(2)^\prime_{HS}$ may be seen as acting on the 2 by 2 matrix $\Phi \equiv (\tilde\phi \,\, \phi)$, where  $\tilde{\phi}\equiv i \tau_2 \phi^*$,  as $\Phi \rightarrow g \, \Phi \, g^\dagger$, 
where $g \in SU(2)'_{HS}$.  The global $SU(2)^\prime_{HS}$ group remains exact in the confined phase because the only Lorentz invariant, $SU(2)_{HS}$ singlet operators one can construct, such as $\phi^\dagger \phi = \tilde \phi^\dagger \tilde \phi$, are invariant under $SU(2)_{HS}^\prime$. In the SM, the global symmetry group $SU(2)^\prime$ is called a ``custodial symmetry''. The name stems from the fact that, although it is explicitely broken by hypercharge and Yukawa interactions, the custodial symmetry  protects the $m_W^2/m_Z^2$ ratio from receiving large radiative corrections. Here the consequence of the $SU(2)^\prime_{HS}$ is to protect the lightest vector boundstates from decaying, so we stick to the name ``custodial symmetry''.

What is the spectrum of this theory? Just like in QCD, where there is nothing like massive gluons, all states must be hidden-color singlets. These come in two categories: bound states of $\phi$ and glueballs. We suppose that the glueballs are more massive than the $\phi$ bound states (like in QCD) and thus we disregard them. The $\phi$ bound states may be organised using the custodial symmetry \cite{'tHooft:1998pk}:
\begin{itemize}
\item $S\equiv \phi^\dagger \phi$ (or equivalently $\tilde{\phi}^\dagger \tilde{\phi}$) bound state, which has spin 0 (S-wave) and is singlet of the custodial symmetry. 
\item $V^-_\mu \equiv \phi^\dagger D_\mu \tilde{\phi}$ and $V^+ _\mu \equiv \tilde{\phi}^\dagger D_\mu \phi$ bound states which are P-waves (spin 1).
\item  $V^0_\mu \equiv (\phi^\dagger D_\mu \phi -\tilde{\phi}^\dagger D_\mu \tilde{\phi})\sqrt{2}$ bound state, also a P-wave state. \end{itemize}
The state $V^0_\mu$ together with $V_\mu^\pm$ form the three components of a triplet under the custodial symmetry.\footnote{
The low energy spectrum is analogous in the Higgs phase of the theory, corresponding to $v_\phi>>\Lambda_{HS}$. Using for simplicity the unitary gauge $\phi=(v_\phi+\eta \,\,, 0)$, at lowest order in $\eta/v_\phi$ all these operators are just the scalar $\eta$ and  $W_{HS}^\pm$ and $Z_{HS}^0$  gauge bosons of $SU(2)_{HS}$, which are respectively  a singlet and the components of a triplet of $SU(2)_{HS}^\prime$ \cite{'tHooft:1998pk}.}

Because of the custodial symmetry, the states in the triplet are degenerate in mass.  Moreover since they are the lightest non-singlet states, all three are stable and thus there is a triplet of DM candidates.\footnote{We assume that the singlet and triplet  are the lightest states in the theory and thus disregard  any, possibly higher, custodial symmetry multiplet  bound states. This appears to be a quite reasonable assumption given the results of lattice simulations of strongly interacting $SU(2)$ theories with scalars in the fundamental repre\-sentation, albeit done in a different context (typically the electroweak theory near the phase transition) \cite{Kajantie:1996mn}.}  In the early universe,  the bound states  form at  the confining phase transition which takes place at  a critical temperature $T_c\sim \Lambda_{HS}$. Since there are no small parameters, the DM and the singlet $S$ are likely to be strongly interacting but their coupling to the SM degrees of freedom takes place only through the perturbative Higgs portal interaction which induces a $S-h-h$ vertex and $S-h$ mixing (${\cal L}_{eff} \sim  \lambda_m \Lambda_{HS} S (v+h)^2$). The hidden and SM sectors are in thermodynamic equilibrium at $T\sim\Lambda_{HS}$, unless the Higgs portal coupling is so small that the two sectors are decoupled, a situation we do not consider here. The relic abundance of DM is then fixed by a Boltzmann equation similar to that used in the standard discussion of the freeze-out of weakly interacting massive particles (WIMP). One difference is that custodial symmetry allows for three vectors interactions, analogous to the three gauge boson vertices in the weakly coupled phase of the theory.  Although we cannot compute the annihilation cross section in the strongly coupled phase of the model, we may distinguish two possible regimes depending on the mass of the singlet state $S$.

I) \underline{The strongly interacting regime (SIMP regime)}. 
If the $S$ bound state is lighter than the DM triplet,\footnote{This is natural on general grounds that higher spin bound states are heavier and is supported by numerical simulations \cite{Kajantie:1996mn}.} DM  should annihilate dominantly into $S S$ pairs, which subsequently decay  (inclusively) into SM states through $S-h$ mixing (like in models of secluded dark matter \cite{Pospelov:2007mp}) or through $S-h-h$ interaction. Since there is no small parameter expansion which holds for these processes, we cannot calculate the annihilation cross section, but we can at least write down the list of annihilations which are allowed by custodial symmetry. At the level of two-body reactions, these are $V_i V_j \leftrightarrow V_k S (h)$ (cross section $\sigma_{ij}$), with $i\neq j\neq k$ in $SU(2)_{HS}^\prime$,  together with $V_i V_i \leftrightarrow S S (h h, Sh)$ ($\sigma_{ii}$). We assume that annihilations into more than two $S$ particles are phase space suppressed. The Boltzmann equation for the abundance of $V$ then takes the form
$$
{d n\over dt} + 3 H n = - {\langle \sigma_{ii} v\rangle\over 3} \left[n^2 - n_{EQ}^2\right] - {\langle\sigma_{ij} v\rangle\over 3} n (n - n_{EQ})\,,
$$
with 
$n = n_1 + n_2 + n_3
$ the density of $V$ states.\footnote{By custodial symmetry $\sigma_{12} = \sigma_{23} = \sigma_{13}$, etc.} The second term is specific to the hidden vector model \cite{Hambye:2008bq} but since $n^2 - n_{EQ}^2 \approx 2 n(n-n_{EQ})$ near freeze-out, the relic abundance behaves as usual $\Omega_{DM} \propto 1/\mbox{\rm Max}(\sigma_{ij}, 2 \sigma_{ii})$. The largest annihilation cross section to a $S$ pair (in s-wave) allowed by unitarity takes the form 
$$
\sigma v_{rel} \approx {4 \pi \over m_{DM}^2 v_{rel}}\,,
$$
with $v_{rel} \approx \sqrt{3/x_f} \sim 1/2.8$ (with $x_f\equiv m_{DM}/T_f\sim 25$ where $T_f$ is the freeze-out temperature).
Since all bound state masses as well as bound state interactions are governed by essentially only one scale, $\Lambda_{HS}\sim m_{DM}$, the resulting annihilation cross section can be written as
\begin{equation}
\sigma_{ii} v_{rel} = \frac{A}{m_{DM}^2}\,,
\end{equation}
with $A \lsim 4 \pi/v_{rel}\sim 35$ from unitarity, but otherwise unknown. 
Taking it between unity and $4 \pi / v_{rel}$ and imposing that the cross section be ${\cal O}(pb)$  \cite{Griest:1989wd} to reach the WMAP abundance $\Omega h^2 \approx 0.11$ puts the DM mass in the range
\begin{equation}
m_{DM} \approx 20-120\, \mbox{\rm  TeV}\,,
\end{equation}
which also sets the scale of $\Lambda_{HS}$.

II) \underline{The weakly interacting regime (WIMP regime)}: For completeness we discuss the (less likely) case $m_S > m_{DM}$. If $S$ is heavier than the V states the DM annihilations to $S$ states is Boltzmann suppressed and eventually it is annihilations into  SM particles, which are controlled by $\lambda_m$, that are dominant. 
Assuming $\Lambda_{HS}>> v$, we may have annihilations into $hh$ pairs or $V h$, with a cross section which is dominated by terms of order $\propto \lambda^2_m/\Lambda_{HS}^2 $ and $\propto \lambda_m^2 v^2/\Lambda_{HS}^4$ respectively. 
The first contribution does not involve any vev because it can proceed directly from the annihilation of 2 DM particles to a $S$ followed by the $\sim \lambda_m  \Lambda_{HS} S h h $ effective interaction, whereas the second which proceeds through an intermediate DM particle necessarily involves $v$ as it has only one $h$ in the final state. 
For small $v/\Lambda_{HS} \ll 1$, the $VV \rightarrow h h$ process should be dominant and   
\begin{equation}
\sigma_{ii} v_{\mbox{\rm rel}} \approx  \lambda_m^2 \frac{B}{m_{DM}^2}\,.
\end{equation}
Taking for the non-perturbative parameter  the same range as above, $1 \lesssim B \lesssim 35$, we can easily get the right relic density, but  possibly for lower DM masses,
\begin{equation}
\label{eq:scenarioII}
m_{DM} \approx (20-120)\,  \lambda_{m}\,\mbox{\rm TeV},
\end{equation}
which can be of order TeV for small enough $\lambda_m$.

\bigskip
In both scenarios, direct detection is purely spin independent, as it proceeds through Higgs exchange. The cross section for direct detection on a nucleus of mass $m_N$ is of the form
\begin{equation}
\sigma_{SI}(V N \rightarrow V N) \sim {\lambda_m^2\over 8\pi} {\Lambda_{HS}^2\over m_{DM}^2 m_S^2}\,{(m_S^2 - m_h^2)^2\over m_S^4 m_h^4} f^2 m_N^2 \mu_r^2\,,
\end{equation}
where $f\sim 0.3$ parameterises the Higgs to nucleon coupling. 
In scenario I, the cross section normalised to one nucleon is $\sigma_{SI} \lesssim \lambda_m^2 10^{-9}$ pb, which, for $\lambda_m$ of order unity,
might  put the candidate  within reach of 100 kg to 1T detectors \cite{Gelmini:2008vi}. In scenario II, $\sigma_{SI} \propto \lambda_m^2/m_{DM}^2$ is independent of $\lambda_m$, using (\ref{eq:scenarioII}), and $\sigma_{SI} \lesssim 10^{-9}$ pb.

Regarding possible indirect detection, scenario I is {\em a priori} the most interesting, as it predicts a heavy DM candidate in the multi TeV range, like that advocated in the recent works that attempt to explain the PAMELA and FERMI excesses \cite{Cirelli:2008pk,Meade:2009iu}. However these scenarios require annihilation of DM into light particles $m \lsim 1$~GeV \cite{ArkaniHamed:2008qn}. In the model discussed in the present paper, we expect that the mass of the $S$  and $V$ states are of the same order of magnitude. Adjusting the bare mass scale $\mu_\phi$ does not help, as we expect both the S-wave and P-wave states to be affected similarly. From this qualitative argument we suspect that  our DM bound state is not  the best  candidate to explain at once all the recent data. 

\bigskip 
A potential issue concerns LEP precision measurements constraints  on mixing between the Higgs and the SM singlet state $S$. These have been discussed in \cite{Hambye:2008bq}
for the case of the Higgs phase of the $SU(2)_{HS}$ gauge group. Given the fact that the $S\--h$ mixing is of order $v/\Lambda_{HS}$, these constraints are easily satisfied in scenario I above, but could be relevant in scenario II.\footnote{Another issue one could raise in presence of strongly self interacting DM, is whether the model satisfy
the various bounds existing on DM self interactions, in particular the bounds of order $\sigma_{self}/m_{DM} \lesssim 1$~cm$^2$~g$^{-1}$ from simulations of large scale structures and cluster profiles \cite{bertone}.
Given that it is the same gauge interactions that induces both the self-interactions and the DM annihilation, resulting in a self-interacting cross section of order the annihilation cross section, and given that DM is very heavy in the above, it can be checked that these bounds are satisfied by many orders of magnitudes.}

Finally, we may entertain the intriguing possibility that there is a relation between the dynamical scale $\Lambda_{HS}$ and the electroweak scale $v$. This could arise if, for some yet to discover fundamental reason, there are no quadratic terms in the full Lagrangian \cite{Patt:2006fw}. If there is confinement in the hidden sector at, say, $\Lambda_{HS} \sim 20$ TeV, the Higgs portal term induces a mass term  in the SM sector
\begin{equation}
\lambda_m \langle \phi^\dagger \phi\rangle H^\dagger H \sim \lambda_m \Lambda_{HS}^2 H^\dagger H\,.
\end{equation}
If $\lambda_m <0$ the Higgs portal induces electroweak symmetry breaking with $v^2 \sim (\lambda_m/\lambda)\, \Lambda_{HS}^2$ and $m_h^2~\sim~\lambda_m \Lambda_{HS}^2$. 
\section*{Acknowledgments}
This work is supported by the FNRS-FRS, the IISN and the Belgian Science Policy (IAP VI-11).

\end{document}